# Charge transport in dual gated bilayer graphene with Corbino geometry

*Jun Yan*[*] *and Michael S. Fuhrer*[†]

Department of Physics, University of Maryland, College Park, MD 20742, USA

Email: [*] juny@umd.edu and [†] mfuhrer@umd.edu

**ABSTRACT** The resistance of dual-gated bilayer graphene is measured as a function of temperature and gating electric fields in the Corbino geometry which precludes edge transport. The temperature-dependent resistance is quantitatively described by a two channel conductance model including parallel thermal activation and variable range hopping channels, which gives the electric-field-dependent band gap whose magnitude is found to be in good agreement with infrared absorption experiments. Low temperature transport is similar to previous studies of dual-gated bilayer graphene with edges, suggesting that edge transport does not play an important role.

The fundamental physics of graphene as well as its potential for future electronic applications has attracted significant contemporary interest.[1-3] The absence of a band gap in single layer graphene (SLG) and bilayer graphene (BLG) poses a significant challenge for many application purposes where the ability to turn off the current is desired.[3] One solution to this problem is the opening of a band gap in BLG by the application of an electric field perpendicular to the graphene layers.[4,5] However, while infrared absorption experiments have observed band gaps up to 250 meV,[6] low temperature electrical transport measurements show that the maximum resistance $R_{max}$ has weak temperature dependence characterized by much smaller energies in either supported[7,8] or suspended[9] BLG with dual-gating. Recently it was suggested that this energy scale difference might result from conducting paths at the edges of BLG samples which are related to the topologically-protected metallic edge states in topological insulators;[10] transport experiments would probe these states while optical experiments would be relatively insensitive to them.

Here we report the first study of dual-gated bilayer graphene in the Corbino disk geometry which removes the possibility of any conductance channel through the edges. The temperature dependence of the conductance is found to be well described by considering a simple thermal activation (STA) channel in parallel with a variable range hopping (VRH) channel. The experimentally extracted band gap magnitude for STA is in good agreement with the values obtained from infrared spectroscopy,[6] and is insensitive to the specific model of variable range hopping used. The low temperature transport behavior, which is consistent with VRH, is qualitatively similar to previous studies with geometries that allow edge transport.[7,8] Our results thus suggest that, at least for these dual-gated BLG devices, the on-off ratio is largely limited by disorder in the bulk BLG rather than by transport along its edges.

Various recent efforts have been made to engineer a band gap in graphene-based field effect transistors. One method is to spatially confine the electrons within a narrow channel by cutting graphene into nanometer size ribbons.[11-14] Another avenue is chemical modification of graphene, such as oxidation, hydrogenation and fluorination.[15-17] While gapped transport behaviors have been revealed in these devices, it has been proven challenging to preserve the superior electrical properties of graphene after such tailoring and modifications. An alternative route is to introduce a sublattice asymmetry in graphene. For SLG, such efforts are so far largely limited to theoretical proposals.[18-20] In the case of BLG, the sublattice asymmetry can be achieved more readily by introducing a potential difference $\delta V$ between the two graphene layers[4,5] as shown in Fig.1(a). Such a potential difference naturally arises in a back-gated BLG device at finite gate fields, resulting in a doped semiconductor. If on the other hand the device is dual gated, the Fermi energy $E_F$ and $\delta V$ can be tuned independently by the difference and sum of the two displacement fields $D_t$ and $D_b$ in the regions above and below the BLG, controlled respectively by the top and bottom gates. Particularly, the condition $D_t = D_b$ results in a charge neutral BLG with $E_F$ lying in a tunable band gap, which opens the opportunity to create an "off" state for the bilayer graphene field effect transistors.[6]

In this manner, recent optical studies on dual-gated BLG convincingly demonstrated a tunable band gap reaching the infrared range.[6] Transport studies on similar dual-gated BLG, however, show that the conductance does not turn off until very low temperatures.[7,8] This greatly limits the on-off ratio and thus the potential usefulness of such devices. Identification of the leakage channel which gives rise to the increased low-temperature conductance is thus not only scientifically interesting but also practically important. Efforts to reduce disorder in BLG such as suspending the device as well as its top gate achieved improvements in $R_{max}$, but the characteristic excitation energy scale at low temperatures is found to be still much smaller than the expected band gap.[9] Another possibility as was proposed recently is the leakage that might occur at the edges of bilayer graphene.[10] Theoretical studies of SLG ribbons show that, depending on its direction with respect to the graphene lattice, the edge could be either conducting or insulating.[21-23] For BLG with zigzag edges, it was shown that there exists bands of edge states inside the electric field induced gap.[24] A recent study found that metallic edge states of gapped BLG may be robust to edge disorder, and could provide a significant leakage conductance path.[10] In addition, asymmetry on the top and bottom gate geometries in real devices would lead to fringing electric fields in the device and a mismatch between the displacement fields $D_t$ and $D_b$, resulting in doping of the edges when the bulk is insulating.

In order to distinguish bulk transport from edge transport, we fabricated dual-gated BLG devices in the Corbino-disk geometry. The BLG sample was deposited on a Si/SiO$_2$ substrate by mechanical exfoliation of natural graphite (Nacional de Grafite Ltda.) and confirmed to be BLG by Raman spectroscopy (see Supporting Information).[25] Figure 1(b) shows a schematic and Figure 1(c) shows the optical image of a dual-gated Corbino-disk BLG device. The ring-shaped top gate is isolated from the BLG by electron-beam overexposed poly-methyl methacrylate (PMMA) (see Supporting Information for more details).[26,27] Note that the device is fabricated such that the source electrode crosses the device from above the top gate electrode, isolated by a second overexposed PMMA layer. This avoids screening of $D_t$ by the source lead which would otherwise result in a leaking path for the device, and ensures an unbroken ring-shaped region of uniform $D_t$ and $D_b$ which separates the source and drain electrodes. The conduction path is thus forced to pass through the bulk of the dual-gated BLG. See supplemental information for details of the device fabrication.

Figure 2(a) shows the room temperature resistance $R$ of the dual-gated BLG Corbino disk as a function of back gate voltage $V_{bg}$ at various values of the top gate voltage $V_{tg}$. The two-peak structure in the data results from two different device regions: the dual-gated region and the regions near the source and drain electrodes which are back-gated only. One resistance peak occurs consistently at $V_{bg}$ = -40V independent of $V_{tg}$ ; we ascribe this to the part of the bilayer graphene that is not covered by the top gate. The other resistance peak moves with top-gate voltage and corresponds to the dual-gated region. This peak occurs at increasingly positive $V_{bg}$ for increasingly negative values of $V_{tg}$ reflecting the condition $D_t = D_b$. This observation offers us a way to estimate the contact resistance which is usually difficult for Corbino geometry devices. We first obtain the resistance of the chromium/gold (Cr/Au) source-drain leads experimentally. In the last step of the sample fabrication, we made two source and two drain Cr/Au leads across the two holes shown in Fig.1(c). We can then measure the source-source or drain-drain resistances bypassing the BLG device; both resistances are approximately 140 Ω. The resistance of the un-top-gated BLG graphene is estimated using the trace with $V_{tg}$ = 30V. At this specific top gate voltage, the two peaks merge into one, indicating that the Fermi energies of the two regions match each other. After subtracting 140 $\Omega$, this trace is then scaled by a numerical factor (0.22) to estimate the un-top-gated bilayer graphene resistance, which can be subtracted to obtain an estimate for the resistance of the dual-gated portion of the device. The inset of Fig.1(a) shows an example of contact resistance subtraction for the $R(V_{bg})$ trace at $V_{tg}$ = -10V.

Figure 2(b) shows the extracted $R(V_{bg})$ data from Fig.2(a) with contact resistance removed. The minimum value of the peak resistance occurs at $V_{tg} \approx$ 15V and $V_{bg} \approx$ 0V, and presumably corresponds to the condition that $D_t = D_b = 0$; here $V_{tg} \neq 0$ indicates that the chemical doping of holes in the dual-gated region predominantly originates from the top surface of BLG, possibly due to the overexposed PMMA. Using the aspect ratio of the top gate, we can further estimate the resistivity of the device as a function of top and back gate voltages. This allows us to obtain the mobility of the device to be about 500 $cm^2V^{-1}s^{-1}$. As a comparison, a previous study reported a dual-gated BLG device with a mobility of 1000 $cm^2V^{-1}s^{-1}$.[7]

Figure 3 shows the resistance of the dual-gated BLG as a function of $V_{tg}$ and $V_{bg}$ at a temperature $T \approx$ 7K. At low temperatures, the top gate has a much larger effect on the peak resistance $R_{max}$; at $V_{tg}$ = -30V, the peak resistivity reaches a value of about 1M$\Omega$. Here the contact resistance is not removed which introduces some error to the resistivity, however since $R_{max}$ at low $T$ at the large displacement fields analyzed below is much larger than the contact resistance, the resulting error to our analysis is quite small. The inset plots the $V_{bg}$ value at which $R_{max}$ occurs as a function of $V_{tg}$. The slope gives the ratio of the two gate capacitances, and indicates that the top gate is about twice as efficient as the back gate. The top gate dielectric thickness is measured to be approximately 130nm by atomic force microscope, from which we estimate the dielectric constant of the overexposed PMMA to be 3.4, somewhat smaller than Ref. 26 but in good agreement with Ref. 27.[28]

We studied the temperature dependence of $R(V_{bg})$ of the dual-gated BLG at several fixed $D_t$'s ranging from 0.78 to 1.17 V/nm.[29] The upper limit of $D_t$ is set by the breakdown electric field of the gate dielectrics, while at low $D_t$'s the smaller temperature dependence of $R_{max}$ would make the errors caused by uncertainties in the contact resistance large. Figure 4 shows $R(V_{bg})$ at various temperatures at $D_t$ =

1.04 V/nm. The inset is the Arrhenius plot for $R_{max}$; linear behavior here would indicate simple thermal activation (STA), i.e. $R_{max} = R_{S0} \exp[\frac{\Delta}{2k_BT}]$ where $\Delta$ is the band gap and $k_B$ is the Boltzmann constant.[8] However, we found that the slope of $\ln R_{max}$ vs. $1/T$ increases monotonically with decreasing $1/T$ (increasing temperature).

Similar changes of the $\ln R_{max}(1/T)$ slope are also observed for other displacement fields as shown in Fig.5(a); $\ln R_{max}$ vs. $1/T$ is always sub-linear in the temperature range studied. Moreover, at about 100K, there is a sharp change in the slope; transport is poorly described by STA at low temperatures. Instead, we find that a plot of $\ln R_{max}$ with respect to $T^{-1/3}$ (Fig.5(b)) shows roughly linear behavior below $T = 100$ K, indicative of variable range hopping (VRH) in a two dimensional system of localized states.[7]

We interpret the full temperature dependence of the resistance as resulting from two conductance channels: VRH dominates at low temperature, and STA to the valence or conduction band edge dominates at high temperature. The conductivities of the two channels are assumed to be additive:

$$\sigma = \sigma_{STA} + \sigma_{VRH}$$
$$\sigma_{STA}^{-1} = R_{S0} \exp\left[\frac{\Delta}{2k_BT}\right] \quad (1)$$
$$\sigma_{VRH}^{-1} = R_{V0} \exp\left[\left(\frac{T_0}{T}\right)^{1/3}\right]$$

The values of $R_{V0}$ and $T_0$ are extracted by fitting the low temperature portion of the data in Fig.5(b). $R_{S0}$ and $\Delta$ are then adjusted to find the best fit to the $R_{max}(T)$ over the whole temperature range.

Figure 5(c) shows the results of the fits of the experimental data using Eqn. (1). The two-channel model explains $R_{max}(T)$ over the entire range of temperatures. Figure 5(d) shows the values of the band gap $\Delta$ obtained from the fits, along with the results obtained in infrared absorption studies in Ref. 6, as well as theoretical calculations.[30,31] The magnitude of the band gap is in good agreement with the optical experiment[6] and the self-consistent tight-binding calculation.[30] Because of the limited temperature range, there is a large uncertainty in determining the power of temperature dependence in the VRH model. Although physically $T^{-1/3}$ is expected for a 2D disordered system, we also tried powers ranging from $T^{-1/4}$ to $T^{-1/2}$ to obtain $R_{V0}$ and $T_0$ before fitting the data over the entire temperature range and found that the values of the band gap $\Delta$ are quite robust and the changes are within the error bars displayed in Fig.5(d). The facts that $\Delta$ does not depend on the details of the low-temperature model used and the magnitude of $\Delta$ is in excellent agreement with optical data together are strong evidence that $\Delta$ measures the band gap in dual-gated BLG and the temperature dependent resistivity indeed reflects STA at high temperatures.

In conclusion, we measured the temperature and displacement field dependent resistance of dual-gated bilayer graphene (BLG) in the Corbino disk geometry. Our data are qualitatively similar to previous results,[7,8] suggesting that for such dual-gated BLG devices, the edges are not dominant conduction paths. The temperature dependence of the resistance is quantitatively explained within a two channel conductance model with high-temperature behavior determined by simple thermal activation, and low-temperature behavior dominated by variable range hopping. The band gap of dual-gated BLG determined from electrical transport is found to be in excellent agreement with infrared spectroscopy studies.[6] The fact that the VRH contribution to the conductance is significant even up to room temperature indicates that reducing disorder in BLG is necessary to increase performance (e.g. on/off ratio) in dual-gated BLG devices.

**Note added.** After submission of this manuscript, we became aware of two related works on dual gated bilayer graphene in Hall bar geometry.[32,33] The similarity of electronic transport between those works and ours for similar temperature range further substantiates our conclusion that edge transport is unimportant in the temperature range we study.

**Acknowledgement.** We thank Amir Yacoby, Alberto F. Morpurgo, Shaffique Adam and Enrico Rossi for discussions. This work is supported by the U.S. ONR MURI, the NSF-UMD-MRSEC Grant No. DMR 05-20471, and an NRI-MRSEC supplement grant.

**Supporting Information Available.** Details on the fabrication of the dual-gated bilayer graphene with Corbino-disk geometry are given, and a Raman spectrum of the bilayer graphene used in this work is shown. This material is available free of charge via the Internet at http://pubs.acs.org.

**FIGURES and FIGURE CAPTIONS**

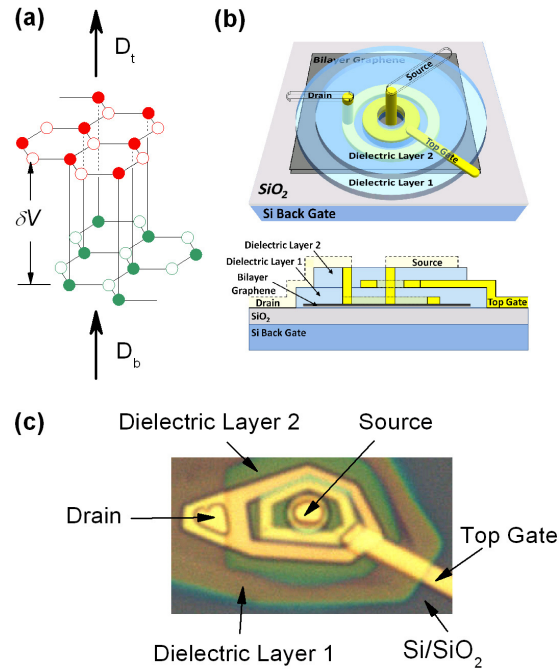

**Figure 1.** Dual-gated bilayer graphene with Corbino-disk geometry. (a) Bilayer graphene lattice sandwiched between top and bottom gating displacement fields. (b) Schematic drawing of the device. The upper plot is the top view and the lower is the side view. The top gate is sandwiched between Dielectric Layer 1 and Dielectric Layer 2. The source lead is on top of Dielectric Layer 2 and contacts the device via the center hole made in the two dielectric layers. (Fabrication details are in the supporting information.) (c) Optical microscope image of a BLG device taken after fabrication of the Source, Drain, and Top Gate electrodes and Dielectric Layers 1 and 2, but before deposition of the source and drain leads. The diameter of the center Cr/Au source electrode is 3μm.

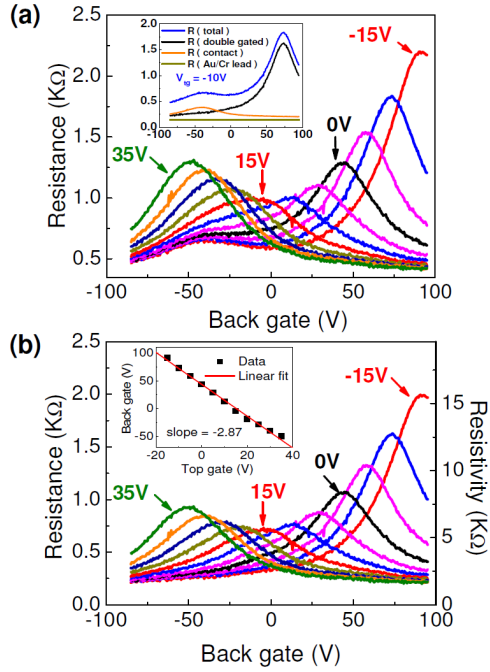

**Figure 2.** Room temperature transport characteristics of the dual-gated Corbino-BLG. (a) Device resistance as a function of back gate voltage $R(V_{bg})$ for various values of $V_{tg}$. $V_{tg}$ is changed in step size of 5V. The inset shows an example of contact resistance subtraction for $V_{tg}$ = -10V. (b) $R(V_{bg})$ after subtraction of contact resistance as described in text. The inset shows the $V_{bg}$ value where the peak resistance occurs for different $V_{tg}$.

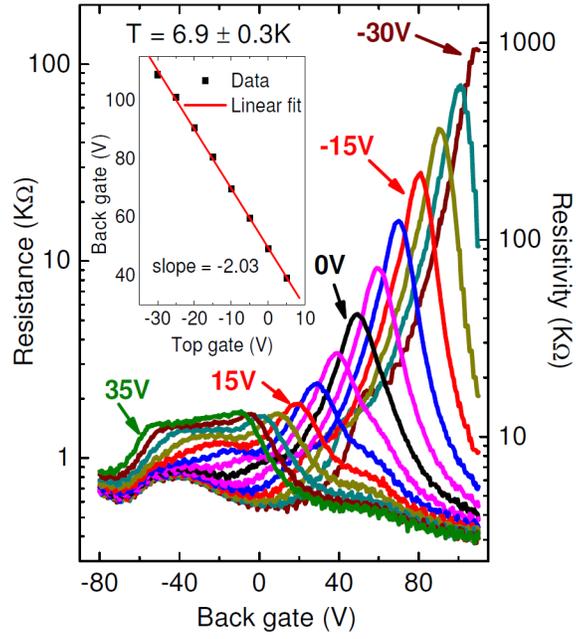

**Figure 3.** Low temperature resistance (left scale) and resisitivity (right scale) of the dual-gated Corbino-BLG as a function of $V_{tg}$ and $V_{bg}$. $V_{tg}$ is changed in step size of 5V. The inset shows the $V_{bg}$ value where the peak resistance occurs for different $V_{tg}$. Here the contact resistance is not removed.

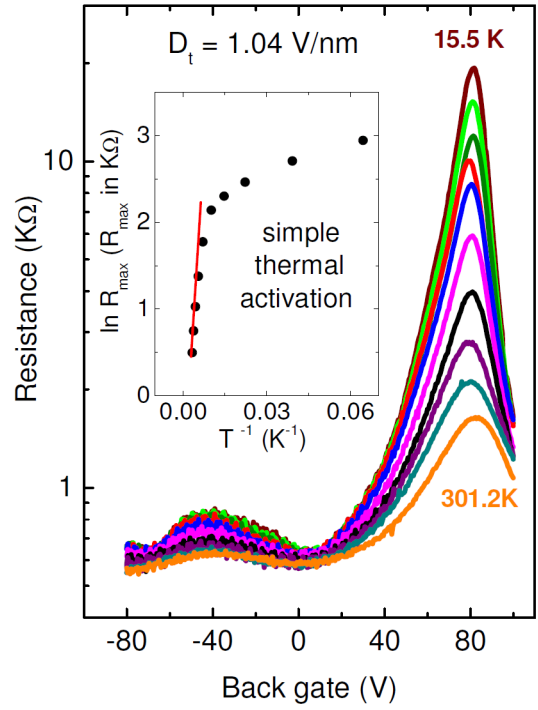

**Figure 4.** Temperature dependence of transport properties of the dual-gated Corbino-BLG at a fixed top gate displacement field of 1.04 V/nm. Different traces top to bottom are for temperatures at 15.5K, 25.4K, 44.6K, 67.5K, 98.5K, 140K, 181K, 222K, 261K and 301.2K. The inset is an Arrhenius plot of the peak resistance; linear behavior corresponds to simple thermal activation.

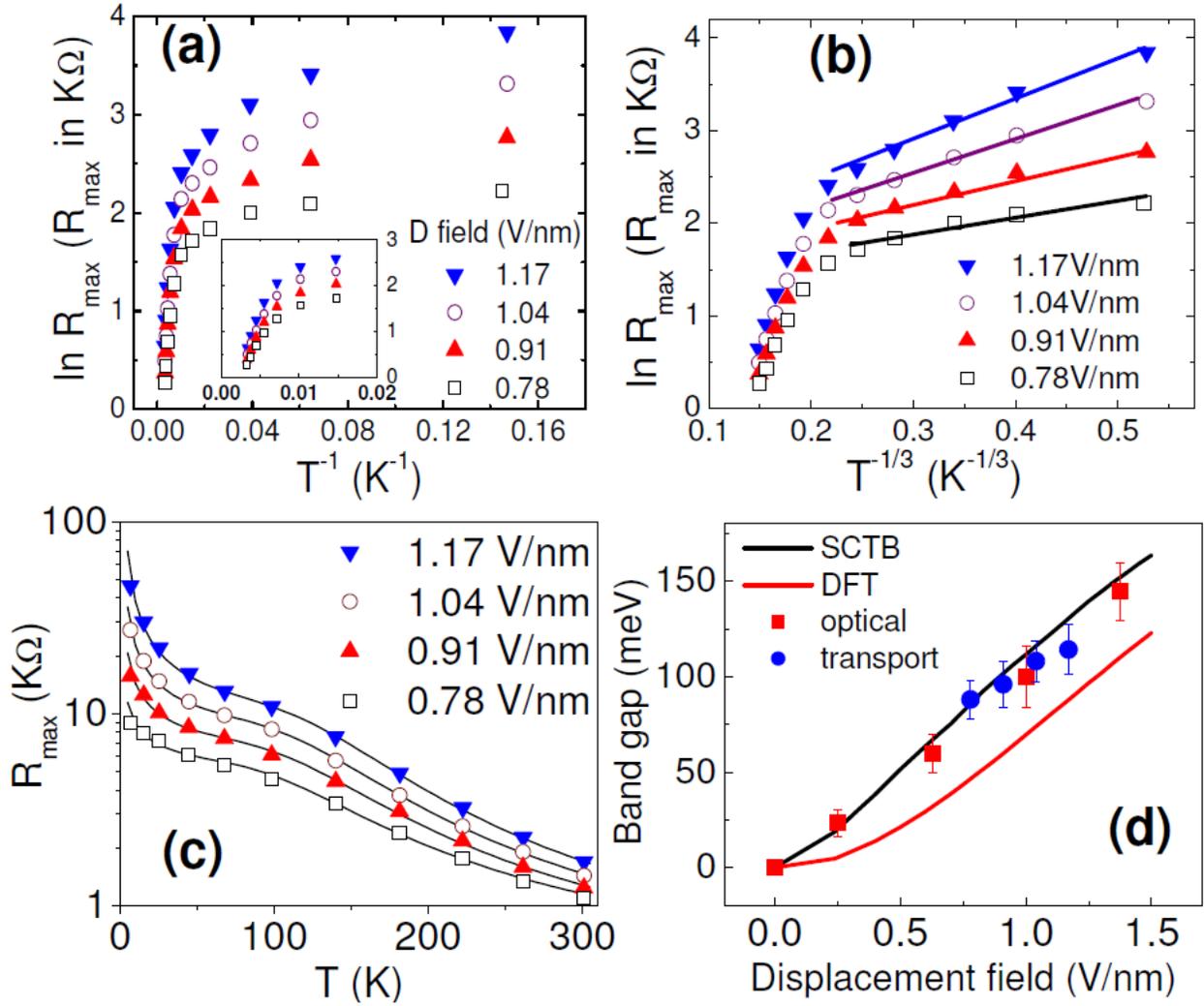

**Figure 5.** Analysis of temperature dependence of the peak resistance $R_{max}(T)$. (a) Arrhenius plot of $R_{max}(T)$ at four different displacement fields. The inset zooms in the high temperature behavior for clarity. (b) Dependence of $\ln R_{max}$ on $T^{-1/3}$; linear behavior indicates variable range hopping in two dimensions. (c) $R_{max}(T)$ with fits to the two channel conductance model (Eqn. (1) in text). Symbols are experimental data and smooth curves are theoretical fits. (d) Band gap of the dual-gated Corbino-BLG. The blue dots are experimental data from our transport measurements. Red squares are from optical studies in Ref. 6. Black and red curves are calculations using self-consistent tight binding (SCTB) model [30] and density functional theories (DFT).[31]